\pdfoutput=1
\documentclass[sigconf]{acmart}

\usepackage{booktabs} 

\usepackage{amsmath}	
\usepackage{amsthm}
\usepackage{amssymb}	
\usepackage{listings}

\usepackage{tikz}
\usetikzlibrary{arrows, decorations, matrix}
\usetikzlibrary{shapes.multipart}
\usetikzlibrary{positioning}
\usetikzlibrary{fit}
\usetikzlibrary{shapes.geometric,backgrounds,calc}

\usepackage{pgfplots}
\usepackage{pgfplotstable}

\pgfplotsset{errorDefaults/.style={
  only marks,
  mark=o,
  mark size=4pt,
  mark options={
    line width=1pt},
  error bars/error bar style={
    line width=1.5pt},
  error bars/error mark options={
    rotate=90,
    mark size=2.5pt,
    line width=1pt},
  error bars/x dir=both,
  error bars/x explicit}}

  \pgfplotsset{symbolDefaults/.style={
    only marks,
    mark size=4pt
  }
  }
\pgfplotsset{ %
  cycle list={ %
    red, mark=*\\%
      blue, mark=square*\\%
      orange, mark=triangle*\\%
      black!40!green, mark=diamond*\\%
  },
}

\usepackage[english]{babel}
\usepackage[T1]{fontenc}
\usepackage[utf8]{inputenc}

\newcommand{\R}{\mathbb{R}}
\newcommand{\N}{\mathbb{N}}
\newcommand{\D}{\mathcal{D}}
\renewcommand{\O}{\mathcal{O}}
\newcommand{\AD}{A\mathcal{D}}
\newcommand{\AO}{A\mathcal{O}}
\newcommand{\eqdot}{\ .}
\newcommand{\aeq}{\kern.35em\text{\small+}\kern-.35em=}
\renewcommand{\d}{\partial}

\DeclareMathOperator{\Arg}{Arg}

\newcount\colveccount
\newcommand*\colvec[1]{
        \global\colveccount#1
        \begin{pmatrix}
        \colvecnext
}
\def\colvecnext#1{
        #1
        \global\advance\colveccount-1
        \ifnum\colveccount>0
                \\
                \expandafter\colvecnext
        \else
                \end{pmatrix}
        \fi
}

\newcount\emptyveccount
\newcommand*\emptyvec[1]{
        \global\emptyveccount#1
        \begin{matrix}
        \emptyvecnext
}
\def\emptyvecnext#1{
        #1
        \global\advance\emptyveccount-1
        \ifnum\emptyveccount>0
                \\
                \expandafter\emptyvecnext
        \else
                \end{matrix}
        \fi
}

\lstdefinestyle{normal}{ %
language=C++,                
numbers=none,
basicstyle=\footnotesize\ttfamily,       
frame=single,                   
tabsize=2,                      
captionpos=b,                   
breaklines=true,                
breakatwhitespace=true,        
}

\lstdefinestyle{numbers}{
language=C++,                
numbers=left,
basicstyle=\footnotesize\ttfamily,       
frame=single,                   
tabsize=2,                      
captionpos=b,                   
breaklines=true,                
breakatwhitespace=true        
}

\lstdefinestyle{xml}{ %
language=xml,                
numbers=none,
basicstyle=\footnotesize\ttfamily,       
frame=single,                   
tabsize=2,                      
captionpos=b,                   
breaklines=true,                
breakatwhitespace=true,        
}

\setcopyright{none}

\acmDOI{???}

\acmISBN{???}

\acmConference[]{}{}{}
\acmYear{}
\copyrightyear{}

\acmArticle{?}
\acmPrice{???}

\editor{???}

\begin{document}
\title{Algorithmic Differentiation for Domain Specific Languages}

\author{Max Sagebaum}
\orcid{1234-5678-9012} 
\affiliation{%
  \institution{Chair for Scientific Computing, TU Kaiserslautern}
  \streetaddress{Paul-Ehrlich-Strasse}
  \city{Kaiserslautern}
  \postcode{67663}
}
\email{max.sagebaum@scicomp.uni-kl.de}

\author{Nicolas R. Gauger}
\orcid{1234-5678-9012} 
\affiliation{%
  \institution{Chair for Scientific Computing, TU Kaiserslautern}
  \streetaddress{Paul-Ehrlich-Strasse}
  \city{Kaiserslautern}
  \postcode{67663}
}

\begin{abstract}
Algorithmic Differentiation (AD) can be used to automate the generation of derivatives in arbitrary software projects.
This will generate maintainable derivatives, that are always consistent with the computation of the software.
If a domain specific language (DSL) is used in a software the state of the art approach is to differentiate the DSL library with the same AD tool.
The drawback of this solution is the reduced performance since the compiler is no longer able to optimize the e.g. SIMD operations.
The new approach in this paper integrates the types and operations of the DSL into the AD tool.
It will be an operator overloading tool that is generated from an abstract definition of a DSL.
This approach enables the compiler to optimize again e.g. for SIMD operation since all calculations are still performed with the original data types.
This will also reduce the required memory for AD since the statements inside the DLS implementation are no longer seen by the AD tool.
The implementation is presented in the paper and first results for the performance of the solution are presented.
\end{abstract}

%
%



\maketitle
\section{Introduction}

Domain Specific Languages (DSLs) \cite{van2000domain} have become increasingly popular for several reasons.
They help to hide domain specific details so that external developers can use the capabilities of the domain without special knowledge.
SIMD operations are a good example for an DSL.
Libraries such as Vc \cite{Kretz2015} provide the capabilities to use the SIMD instructions of an arbitrary processor architecture, the user can focus on writing the application code.

DSLs for SIMD optimization are often used in software on HPC clusters.
An example can be the simulation of the fluid dynamics around a wing body.
If this wing body needs to be modified such that e.g. the drag is reduced a shape optimization is necessary.
In order to perform the shape optimization accurate sensitivity information needs to be available.
A general concept, for the semi automatic generation of sensitivity information, is the application of Algorithmic Differentiation (AD) \cite{grie08} to the simulation software.
AD describes how a computer program can be interpreted as the concatenation of several millions of elemental operators like $+$, $*$, $sin$ and $exp$.
The chain rule and the directional derivative is applied on this large concatenation of functions.
This yields the forward mode of AD.
The reverse mode of AD is introduced by applying the discrete adjoint calculus \cite{dunford1958linear,apostol1969calculus} on the forward mode.

The reverse mode of AD has to store a certain set of information for each elemental operation that is evaluated.
If this is an operation of an DSL there are two concepts how this operation can be treated.
Traditional operator overloading AD  tools like ADOL-c \cite{walther2009getting} or CoDiPack \cite{sagebaum2017high} insert their AD type into the library for the DSL.
Every operation inside the library is then evaluated with the type from the AD tool.
For a simple linear system solve with a shifted right hand side
\begin{equation*}
	r = M^{-1} (v_2 - v_1) + v_1
\end{equation*}
these tools would store the information for several thousands of elemental operations.
An example implementation of the above equation in CoDiPack requires 62000 bytes for a vector dimension of 10.

The second concept is to treat the elemental operation as an external function and provide specialized code for the sensitivity computation.
This process is quite involved, error prone and has to be repeated for every new application.
An external function introduces an overhead in time and memory which makes it only feasible for, in this case, large linear systems.
For small dimensions it would not be efficient.

The approach proposed in this paper adds the functionality of a DSL directly to the set of elemental functions that the AD tool can handle.
This eliminates the overhead from the external function approach and does not need information from intermediate statements.
If this is done for the above example the required memory would be 1200 bytes which is a reduction by the factor of 51.

However, difficulties arise when implementing this idea.
Instead of handling just the floating point type of the programming language, the AD tool needs to handle all types introduced by the DSL.
Furthermore, the implementation of the elemental functions is problematic.
There are several thousands of DLSs already available and an AD developer can not handle all of them.
The AD developer typically has not the knowledge about the DSL to implement the sensitivity computation.
On the other hand, the developer of the DSL has not the knowledge to implement an AD tool for his language.
Therefore, we want to develop a generalized AD tool in this paper.
The developer of the DSL has to provide the specification for the types and operations of the DSL, and the derivative computation for each operation.
The generalized AD tool will then be generated with this information and can be used to compute sensitivities.

This paper will first give an introduction to AD and explain very briefly how operator overloading tools are implemented.
Afterwards the concept of AD for DSL is explained in more detail.
The design goals and implementation details for the generalized AD tool are shown in the following chapters.
The performance of the implementation is compared using examples with and without DSLs.

\section{Algorithmic Differentiation}
\subsection{Basic theory}
\label{sec.basicAD}
In this paper we are only providing a very brief introduction to AD, for a complete derivation of the reverse AD mode see \cite{grie08, naumann2012art}. AD will be applied on the function
\begin{equation}
	y = F(x)
\end{equation}
with $x \in \R^m$ and $y \in \R^n$.
The reverse AD mode for this equation is, following the derivation in \cite{grie08},
\begin{equation}
	\bar x \aeq \frac{\d F}{\d x}^T(x) \bar y
\end{equation}
with $\bar y \in \R^n$ and $\bar x \in \R^m$ and describes what the reverse AD mode computes.
The adjoint solution $\bar x$ is computed in the background by evaluating
\begin{equation}
\label{eq.eleBack}
\bar u \aeq \frac{\d \phi}{\d u}^T(u) \bar w; \quad \bar w = 0
\end{equation}
with $u \in \R^l$, $w \in \R$, $\bar w \in \R$ and $\bar u \in \R^l$ for every elemental operation $\phi$ that computes $F$.
$F$ can consist of several millions of elemental operations $\phi_i: \R^{l_i} \rightarrow \R$ with $l_i \in \N$ and $i \in \N$ which are evaluated from $1$ to $K$.
In order to evaluate equation \eqref{eq.eleBack}, AD stores information for every $\phi_i$ during the primal evaluation.
Afterwards, AD evaluates equation \eqref{eq.eleBack} for $i$ from $K$ to $1$ with the stored information.
The sensitivity information is propagated in reverse order from the output variables to the input variables.
An example elemental function can be $h(a, b) = a \cdot b$ with $l=2$.
The reverse AD evaluation would then be
\begin{equation}
	\colvec{2}{\bar a}{\bar b} \aeq \colvec{2}{b}{a} \bar w \eqdot
\end{equation}

\subsection{Theory for extended operators}
The definition of $\phi$ in \eqref{eq.eleBack} assumes that an elemental function can only have one output element.
This restriction can be lifted as described in \cite{grie08}.
Let $\phi_{i \ldots j} : \R^l \rightarrow \R^k$ be a function with $k \in \N$ output variables with $k = j - i + 1$.
$\phi_{i \ldots j}$ can be written as $\phi_{i \ldots j}(u) = (\phi_{i}(u), \phi_{i + 1}(u), \ldots, \phi_{j - 1}(u), \phi_{j}(u))$.
$\phi_i$ to $\phi_j$ are elemental operators in the context of AD and can be treated with the theory presented above.
The only difference is the special dependencies of the operators $\phi_i$ to $\phi_j$ which can be exploited in the implementation for domain specific languages.

\section{Operator overloading AD implementation}
AD can be applied to a source code in two ways.
The source transformation approach generates a new source code that adds  additional statements for the reverse AD mode.
This approach is mostly applied via Tapenade \cite{TapenadeRef13} or ADIC \cite{Bischof1996AAD} to Fortran codes.
The second approach introduces AD through the operator overloading capabilities of a language like C++.
All operators for a new computation class are overloaded and store the required information for the reverse AD mode in the background.
After the program is evaluated the stored information is interpreted in the reverse order and the sensitivities are propagated from the output values of the program to the input values.
For a detailed description of an AD tool implementation see \cite{sagebaum2017high, Hogan2014FRM}.

In general each AD tool implements a \emph{tape} structure where the information is stored.
The tape consists of several global data vectors and data streams.
In this paper we are mainly considering a primal value taping approach.
Here, the input values $u$ in equation \eqref{eq.eleBack} are stored.
For the adjoint variable identification (e.g. $\bar u$) a reuse index management scheme is used.
This means that the identifier of a destructed variable can be assigned to a different variable after the destruction.
A primal value taping approach with a reuse index management scheme requires two global vectors.
The first one is the primal value vector and holds the primal values of all variables that are currently used in the program.
The second vector is the adjoint vector which holds all corresponding adjoint values for each primal value (e.g. $\bar u$ for $u$).
These vectors are accessed in a random access pattern and are stored in random access memory (RAM).
The taping scheme requires furthermore six data stream.
For each elemental operator (statement) the identifier of the output value (left hand side (lhs)), the old value of the left hand side, the function handle for the evaluation of equation \eqref{eq.eleBack} and the number of active arguments are required.
For each argument of an elemental operator (right hand side (rhs)) the identifier is required.
The sixth data stream consists of the data for all constant arguments.
For details on the required data see \cite{sagebaum2018Expr}.
Figure \ref{fig.genTape} shows the data for the taping scheme in a graphical way and highlights the required data for a sample elemental operation $h$.
It also shows the index manager which holds all indices that are currently not used.
The index manager can be queried for new indices where $i_w$ describes the index that corresponds to $w$ and can be used to access the adjoint $\bar w$.

\colorlet{colA}{black!30!yellow}
\colorlet{colB}{black!20!red}
\colorlet{colC}{black!40!orange}
\colorlet{colD}{black!40!blue}
\definecolor{colW}{HTML}{40bf40}
\definecolor{colH}{HTML}{ac00e6}
\colorlet{colCons}{black!60!green}

\def\mathunderline#1#2{\color{#1}\underline{{\color{black}#2}}\color{black}}
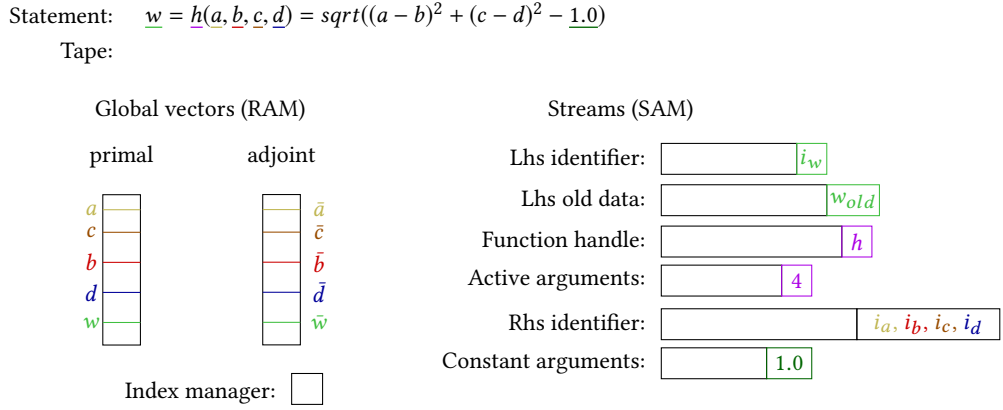
\begin{figure*}
	\begin{tikzpicture}[node distance=1cm]
	  \tikzset{
	    fitting node/.style={
	      inner sep=0pt,
	      fill=none,
	      draw=none,
	      reset transform,
	      fit={(\pgf@pathminx,\pgf@pathminy) (\pgf@pathmaxx,\pgf@pathmaxy)}
	    },
	    reset transform/.code={\pgftransformreset}
	  }
	  \tikzset{
	      mymark/.style={},
	      myarrow/.style={->, >=latex', shorten >=1pt, thick},
	      myarrow2/.style={->, >=latex', shorten >=1pt, thick, dashed},
	      mylabel/.style={text width=7em, text centered}
	  }
	  \newcommand\IndexColor{blue}
	  
	  \node[anchor=east](stmtName) {Statement:};
	  \node[right=0.2cm of stmtName, anchor=west](stmt) { $\mathunderline{colW}{w} = \mathunderline{colH}{h}(\mathunderline{colA}{a},\mathunderline{colB}{b},\mathunderline{colC}{c},\mathunderline{colD}{d}) = sqrt((a-b)^2 + (c-d)^2 - \mathunderline{colCons}{1.0})$};
	  
	  \node[below=0.5cm of stmtName.east, anchor=east] (tapeName) {Tape:};
	  \node[below=0.5cm of tapeName, anchor=west] (vectorsName) {Global vectors (RAM)};
	  \node[below=0.25cm of vectorsName] (belowVec) {};
	  \node[left=0.4cm of belowVec] (primalName) {primal};
	  \node[right=0.4cm of belowVec] (adjointName) {adjoint};
	  
	  \draw[black] (primalName.south) ++(-0.25cm, -0.25cm) rectangle ++(0.5cm, -2cm);
	  \draw[black] (adjointName.south) ++(-0.25cm, -0.25cm) rectangle ++(0.5cm, -2cm);

	  \draw[colA] (primalName.south) ++(0.25cm, -0.45cm) -- ++(-0.5cm, 0.0cm) ++ (-0.15cm , 0.0cm) node {$a$};
	  \draw[colC] (primalName.south) ++(0.25cm, -0.75cm) -- ++(-0.5cm, 0.0cm) ++ (-0.15cm , 0.0cm) node {$c$};
	  \draw[colB] (primalName.south) ++(0.25cm, -1.15cm) -- ++(-0.5cm, 0.0cm) ++ (-0.15cm , 0.0cm) node {$b$};
	  \draw[colD] (primalName.south) ++(0.25cm, -1.55cm) -- ++(-0.5cm, 0.0cm) ++ (-0.15cm , 0.0cm) node {$d$};
	  \draw[colW] (primalName.south) ++(0.25cm, -1.95cm) -- ++(-0.5cm, 0.0cm) ++ (-0.15cm , 0.0cm) node {$w$};
	  
	  \draw[colA] (adjointName.south) ++(0.25cm, -0.45cm) -- ++(-0.5cm, 0.0cm) ++ (0.75cm , 0.0cm) node {$\bar a$};
	  \draw[colC] (adjointName.south) ++(0.25cm, -0.75cm) -- ++(-0.5cm, 0.0cm) ++ (0.75cm , 0.0cm) node {$\bar c$};
	  \draw[colB] (adjointName.south) ++(0.25cm, -1.15cm) -- ++(-0.5cm, 0.0cm) ++ (0.75cm , 0.0cm) node {$\bar b$};
	  \draw[colD] (adjointName.south) ++(0.25cm, -1.55cm) -- ++(-0.5cm, 0.0cm) ++ (0.75cm , 0.0cm) node {$\bar d$};
	  \draw[colW] (adjointName.south) ++(0.25cm, -1.95cm) -- ++(-0.5cm, 0.0cm) ++ (0.75cm , 0.0cm) node {$\bar w$};
	  
	  \node[right=3cm of vectorsName] (streamsName) {Streams (SAM)};
	  
	  \node[below=0.25cm of streamsName] (belowStream) {};
	  \node[left=-0.55cm of belowStream, anchor=east] (1StreamName) {Lhs identifier:};
	  \node[below=0.55cm of 1StreamName.east, anchor=east] (2StreamName) {Lhs old data:};
	  \node[below=0.55cm of 2StreamName.east, anchor=east] (3StreamName) {Function handle:};
	  \node[below=0.55cm of 3StreamName.east, anchor=east] (4StreamName) {Active arguments:};
	  \node[below=0.55cm of 4StreamName.east, anchor=east] (5StreamName) {Rhs identifier:};
	  \node[below=0.55cm of 5StreamName.east, anchor=east] (6StreamName) {Constant arguments:};
	  
	  \draw[black] (1StreamName.north east) ++(0.1cm, -0.05cm) rectangle ++(1.8cm, -1.3em);
	  \draw[black] (2StreamName.north east) ++(0.1cm, -0.05cm) rectangle ++(2.2cm, -1.3em);
	  \draw[black] (3StreamName.north east) ++(0.1cm, -0.05cm) rectangle ++(2.4cm, -1.3em);
	  \draw[black] (4StreamName.north east) ++(0.1cm, -0.05cm) rectangle ++(1.6cm, -1.3em);
	  \draw[black] (5StreamName.north east) ++(0.1cm, -0.05cm) rectangle ++(2.6cm, -1.3em);
	  \draw[black] (6StreamName.north east) ++(0.1cm, -0.05cm) rectangle ++(1.4cm, -1.3em);
	  
	  \draw[colW] (1StreamName.north east) ++(1.9cm, -0.05cm) rectangle ++(0.4cm, -1.3em) node[pos=0.5] {$i_w$};
	  \draw[colW] (2StreamName.north east) ++(2.3cm, -0.05cm) rectangle ++(0.7cm, -1.3em) node[pos=0.5] {$w_{old}$};
	  \draw[colH] (3StreamName.north east) ++(2.5cm, -0.05cm) rectangle ++(0.4cm, -1.3em) node[pos=0.5] {$h$};
	  \draw[colH] (4StreamName.north east) ++(1.7cm, -0.05cm) rectangle ++(0.4cm, -1.3em) node[pos=0.5] {$4$};
	  \draw[black] (5StreamName.north east) ++(2.7cm, -0.05cm) rectangle ++(1.9cm, -1.3em) node[pos=0.5] {\color{colA}$i_a$, \color{colB}$i_b$, \color{colC}$i_c$, \color{colD}$i_d$};
	  \draw[colCons] (6StreamName.north east) ++(1.5cm, -0.05cm) rectangle ++(0.6cm, -1.3em) node[pos=0.5] {$1.0$};
	  
	  \node[below=3.25cm of vectorsName] (indexManagerName) {Index manager:};
	  \draw[black] (indexManagerName.north east) ++(0.1cm, 0.0cm) rectangle ++(1.3em, -1.3em);

	\end{tikzpicture}

	\caption{Data of an AD tape for primal value taping with index reuse.}
	\label{fig.genTape}
\end{figure*}
\section{AD for domain specific languages}

Each AD tool provides an AD type that is then used in the application.
Here we assume that the type is called \emph{Real}.
A general use would then be:
\begin{lstlisting}
Real w = sqrt(pow(a-b, 2) + pow(c-d, 2) - 1.0);
\end{lstlisting}
Any DSL can be differentiated with AD by exchanging the calculation type of the implementation.
If the DSL is templated this can be archived by changing the template argument.
That is, \lstinline|Vector<double>| would become \lstinline|Vector<Real>|.
The introduction showed that this approach can be rather inefficient since all intermediate steps are recorded by the AD tool.
It can also not be applied if the DSL is defined in language intrinsics like the \lstinline|F64vec4| type for a four element double vector from the AVX instruction set \cite{firasta2008intel}.

Lets assume that the DSL is named $L$ and consists of a finite set of data objects $D \in \D$ and a finite set of operations $o \in \O$ where $\Arg(o)$ describes the ordered set of arguments for $o$ with $D \in \D$ for all $D$ in $\Arg(o)$.
The language is then specified via $L(\D, \O)$.
The traditional approach would now modify the objects and operations by using the AD type for the implementation.
This generates the language $L_{\text{Real}}(\D_{\text{Real}}, \O_{\text{Real}})$.
The new approach of this paper takes all data objects $D \in \D$ and creates a new set of objects $\AD = \{ \{D, id_D \} \ |\  \forall D \in \D \} \cup \D$ where $id_D$ is an identifier for a data object of type $D$.
For each operator a new set of  operators is created by changing the arguments of the operator to the corresponding data objects in $\AD$.
This is done for each subset of arguments such that $\AD$ data objects and $\D$ data objects are intermixed.
The mathematical formulation is now
\begin{equation*}
	\AO = \{ \pi(o) \ | \  \forall o \in \O, \forall i = 1, \ldots, |\Arg(o)|, \forall \pi \in \colvec{2}{\Arg(o)}{i}\}
\end{equation*}
where $\pi \in \colvec{2}{\Arg(o)}{i}$ describes a subset with $i$ elements.
The set of the arguments is then defined as
\begin{align*}
	\Arg(\pi(o)) := & \{ \{D, id_D\} \ |\ \forall D \in \Arg(o) \text{ s.t. } D \in \pi(o)\} \\
	& \cup \{ D \ |\ \forall D \in \Arg(o) \text{ s.t. } D \not\in \pi(o)\} \eqdot
\end{align*}

The new differentiated language is defined as $AL(\AD, \AO)$ and is based on the language $L(\D, \O)$.
In terms of software engineering the old process can be described as defining the type Real as
\begin{lstlisting}
	struct Real {
	    double p;
	    int id;
	};
\end{lstlisting}
and transforming the language by changing \lstinline|Vector<double>| to \lstinline|Vector<Real>|.
The new process defines the AD type
\begin{lstlisting}
	template< typename T>
	struct DSLReal {
	    T p;
	    int id;
	};
\end{lstlisting}
and then creates a new language by using \lstinline|DSLReal<Vector<double> >|.
That is, instead of putting the AD type inside the structure, the AD type is wrapped around the structure.

\section{Designing AD for domain specific languages}
The implementation of the new language $AL$ from the language $L$ should not be done for only one sample DSL.
There are several thousand DSLs available and new ones are created regularly.
An AD implementation for DSLs needs to have the following properties:
\begin{itemize}
	\item Independent of the DSL
	\item Easy to adapt to new DSLs
	\item No AD implementation knowledge of the DSL developer/user required 
\end{itemize}
The first property is already motivated above.
The second one describes the general usability of the implementation.
For a library developer it should be quite simple to describe his DSL to the AD tool implementation.
The third property is important to guarantee the usability of the tool.
The developer of an DSL, who wants to provide an AD capability of his language, should only provide the derivative information and nothing more.
The AD specific data management is then handled by the AD tool which is presented in this paper.

The global vectors from figure \ref{fig.genTape} need to be addressed first.
If the design is left as it is then all data objects of the language $L$ need to be converted into the data format of this vector which can introduce a significant overhead.
A further problem is the memory alignment of the data.
SIMD types for example should not be instantiated at arbitrary memory locations.
The design of one global vector is therefore droped.
Each data object will have its own global vectors which can then be configured for the requirements of the data object.
Since the index manager translates identifiers into positions of the global vectors, there needs to be one index manager for each data object.

\begin{figure*}
	\begin{tikzpicture}[node distance=1cm]
	  \tikzset{
	    fitting node/.style={
	      inner sep=0pt,
	      fill=none,
	      draw=none,
	      reset transform,
	      fit={(\pgf@pathminx,\pgf@pathminy) (\pgf@pathmaxx,\pgf@pathmaxy)}
	    },
	    reset transform/.code={\pgftransformreset}
	  }
	  \tikzset{
	      mymark/.style={},
	      myarrow/.style={->, >=latex', shorten >=1pt, thick},
	      myarrow2/.style={->, >=latex', shorten >=1pt, thick, dashed},
	      mylabel/.style={text width=7em, text centered}
	  }
	  \newcommand\IndexColor{blue}
	  
	  \node[anchor=east](stmtName) {Statement:};
	  \node[right=0.2cm of stmtName, anchor=west](stmt) { $\mathunderline{colW}{w} = \mathunderline{colH}{h}(\mathunderline{colA}{A} \in \R^{3 \times 3},\mathunderline{colD}{D} \in \R^{3 \times 3},\mathunderline{colB}{b} \in \R^3,\mathunderline{colC}{c} \in \R^3) = (\mathunderline{colCons}{1.0} - D + A).solve(c - b)$};
	  
	  \node[below=0.5cm of stmtName.east, anchor=east] (tapeName) {Tape:};
	  \node[below=0.5cm of tapeName, anchor=west] (belowTape) {};
	  	  
	  \node[left=3.5cm of belowTape, anchor=west] (vectorsName) {Global vectors (RAM): Matrix};
	  \node[below=0.25cm of vectorsName] (belowVec) {};
	  \node[left=0.0cm of belowVec] (primalName) {primal};
	  \node[right=0.0cm of belowVec] (adjointName) {adjoint};
	  
	  \draw[black] (primalName.south) ++(-0.25cm, -0.25cm) rectangle ++(0.5cm, -2cm);
	  \draw[black] (adjointName.south) ++(-0.25cm, -0.25cm) rectangle ++(0.5cm, -2cm);

	  \draw[colA] (primalName.south) ++(0.25cm, -0.45cm) -- ++(-0.5cm, 0.0cm) ++ (-0.15cm , 0.0cm) node {$A$};
	  \draw[colD] (primalName.south) ++(0.25cm, -1.55cm) -- ++(-0.5cm, 0.0cm) ++ (-0.15cm , 0.0cm) node {$D$};
	  
	  \draw[colA] (adjointName.south) ++(0.25cm, -0.45cm) -- ++(-0.5cm, 0.0cm) ++ (0.75cm , 0.0cm) node {$\bar A$};
	  \draw[colD] (adjointName.south) ++(0.25cm, -1.55cm) -- ++(-0.5cm, 0.0cm) ++ (0.75cm , 0.0cm) node {$\bar D$};
	  
	  \node[below=3.25cm of vectorsName] (indexManagerName) {Index manager:};
	  \draw[black] (indexManagerName.north east) ++(0.1cm, 0.0cm) rectangle ++(1.3em, -1.3em);
	  
	  \node[draw,black, fit=(vectorsName) (indexManagerName)] (regionMatrix) {};
	  
	  \node[right=0.25cm of vectorsName, anchor=west] (vectorsName) {Global vectors (RAM): Vector};
	  \node[below=0.25cm of vectorsName] (belowVec) {};
	  \node[left=0.0cm of belowVec] (primalName) {primal};
	  \node[right=0.0cm of belowVec] (adjointName) {adjoint};
	  
	  \draw[black] (primalName.south) ++(-0.25cm, -0.25cm) rectangle ++(0.5cm, -2cm);
	  \draw[black] (adjointName.south) ++(-0.25cm, -0.25cm) rectangle ++(0.5cm, -2cm);

	  \draw[colC] (primalName.south) ++(0.25cm, -0.75cm) -- ++(-0.5cm, 0.0cm) ++ (-0.15cm , 0.0cm) node {$c$};
	  \draw[colB] (primalName.south) ++(0.25cm, -1.15cm) -- ++(-0.5cm, 0.0cm) ++ (-0.15cm , 0.0cm) node {$b$};
	  \draw[colW] (primalName.south) ++(0.25cm, -1.95cm) -- ++(-0.5cm, 0.0cm) ++ (-0.15cm , 0.0cm) node {$w$};
	  
	  \draw[colC] (adjointName.south) ++(0.25cm, -0.75cm) -- ++(-0.5cm, 0.0cm) ++ (0.75cm , 0.0cm) node {$\bar c$};
	  \draw[colB] (adjointName.south) ++(0.25cm, -1.15cm) -- ++(-0.5cm, 0.0cm) ++ (0.75cm , 0.0cm) node {$\bar b$};
	  \draw[colW] (adjointName.south) ++(0.25cm, -1.95cm) -- ++(-0.5cm, 0.0cm) ++ (0.75cm , 0.0cm) node {$\bar w$};
	  
	  \node[below=3.25cm of vectorsName] (indexManagerName) {Index manager:};
	  \draw[black] (indexManagerName.north east) ++(0.1cm, 0.0cm) rectangle ++(1.3em, -1.3em);
	  
	  \node[draw,black, fit=(vectorsName) (indexManagerName)] (regionVector) {};
	  
	  \node[draw,black, right=0.05cm of regionVector, minimum height=13.9em] (regionOther) {$\ldots$};
	  
	  \node[right=2.5cm of vectorsName] (streamsName) {Streams (SAM)};
	  
	  \node[below=0.25cm of streamsName] (belowStream) {};
	  \node[left=-0.55cm of belowStream, anchor=east] (1StreamName) {Lhs identifier:};
	  \node[below=0.55cm of 1StreamName.east, anchor=east] (2StreamName) {Lhs old data:};
	  \node[below=0.55cm of 2StreamName.east, anchor=east] (3StreamName) {Function handle:};
	  \node[below=0.55cm of 3StreamName.east, anchor=east] (4StreamName) {Active arguments:};
	  \node[below=0.55cm of 4StreamName.east, anchor=east] (5StreamName) {Rhs identifier:};
	  \node[below=0.55cm of 5StreamName.east, anchor=east] (6StreamName) {Constant arguments:};
	  
	  \draw[black] (1StreamName.north east) ++(0.1cm, -0.05cm) rectangle ++(1.8cm, -1.3em);
	  \draw[black] (2StreamName.north east) ++(0.1cm, -0.05cm) rectangle ++(2.2cm, -1.3em);
	  \draw[black] (3StreamName.north east) ++(0.1cm, -0.05cm) rectangle ++(2.4cm, -1.3em);
	  \draw[black] (4StreamName.north east) ++(0.1cm, -0.05cm) rectangle ++(1.6cm, -1.3em);
	  \draw[black] (5StreamName.north east) ++(0.1cm, -0.05cm) rectangle ++(2.6cm, -1.3em);
	  \draw[black] (6StreamName.north east) ++(0.1cm, -0.05cm) rectangle ++(1.4cm, -1.3em);
	  
	  \draw[colW] (1StreamName.north east) ++(1.9cm, -0.05cm) rectangle ++(0.4cm, -1.3em) node[pos=0.5] {$i_w$};
	  \draw[colW] (2StreamName.north east) ++(2.3cm, -0.05cm) rectangle ++(0.7cm, -1.3em) node[pos=0.5] {$w_{old}$};
	  \draw[colH] (3StreamName.north east) ++(2.5cm, -0.05cm) rectangle ++(0.4cm, -1.3em) node[pos=0.5] {$h$};
	  \draw[colH] (4StreamName.north east) ++(1.7cm, -0.05cm) rectangle ++(0.4cm, -1.3em) node[pos=0.5] {$4$};
	  \draw[black] (5StreamName.north east) ++(2.7cm, -0.05cm) rectangle ++(1.9cm, -1.3em) node[pos=0.5] {\color{colA}$i_A$, \color{colB}$i_b$, \color{colC}$i_c$, \color{colD}$i_D$} node[pos=1.0] (tempNode) {};
	  \draw[colCons] (6StreamName.north east) ++(1.5cm, -0.05cm) rectangle ++(0.6cm, -1.3em) node[pos=0.5] {$1.0$};
	  
	  \node[draw,black, fit=(streamsName) (6StreamName) (tempNode)] (regionStreams) {};

	\end{tikzpicture}

	\caption{Data of an AD tape for domain specific languages. Storing method is primal value taping with index reuse.}
	\label{fig.specTape}
\end{figure*}
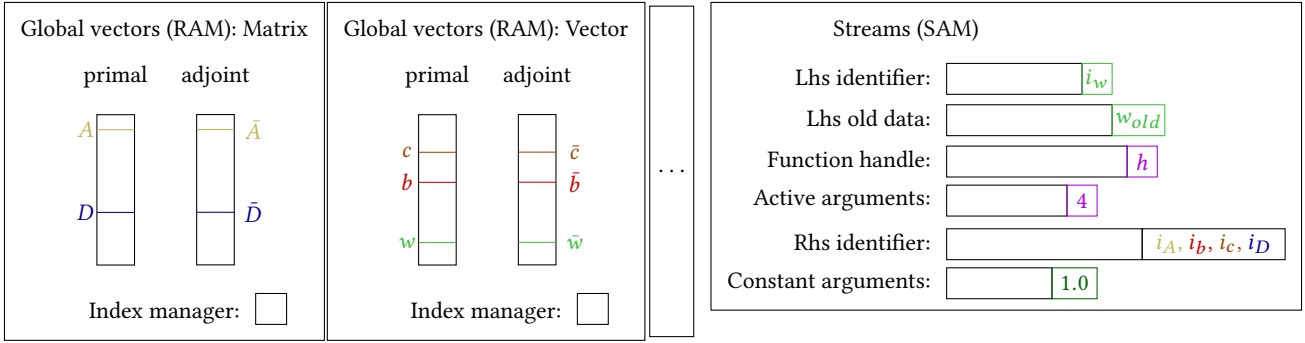

The second design decision concerns the data streams of the tape in figure \ref{fig.genTape}.
From the six required streams the left hand side old data and the constant value data stream have to hold object specific data.
If the same approach is used as for the global vectors it would introduce two data stream for each data object of the language $L$.
The synchronization and management is then much more involved and can currently not be handled with the library for the stream management.
Therefore, the stream structure is not changed and all data objects need to be converted to and from the streams.
These two decisions lead to the new tape layout in figure \ref{fig.specTape}.
The design consists now of the global tape and secondary tapes that hold the information for each data object.

This covers the basic design of the new tape for the language $AL$.
It is now important to specify the information for the reverse evaluation.
The reverse AD equation \eqref{eq.eleBack} from section \ref{sec.basicAD} is reformulated in the context of the language $AL$.
Let $o$ be in $\AO$ then the primal evaluation is
\begin{equation}
	w = o(u_1, \ldots, u_l)
\end{equation}
with $w \in \AD$ and $u_i \in \AD$.
$o$ takes in this case the place of the elemental operator $\phi$.
The reverse AD equation can now be formulated in the same context as
\begin{equation}
	\bar u_i \aeq \frac{\d o}{\d u_i}(u_1, \ldots, u_l)^T \bar w \quad \forall i = 1, \ldots, l; \bar w = 0 \eqdot
\end{equation}

The implementation of the AD tool will provide all values for the inputs $u_i$, the output $w$ and the adjoint variables.
The missing information is the computation of the derivatives $\frac{\d o}{\d u_i}$.
Since this is a DSL specific information, the user needs to provide it.
If only the derivative $\frac{\d o}{\d u_i}$ is made available the type of the object is not clear.
If the computation of $\frac{\d o}{\d u_i}^T \bar w$ is provided the result is a data object of the same type as $u_i$ and the computation might be implemented in an optimized way.
Therefore, the third design decision is that the user has to provide functions for the computation of $o(u_1, \ldots, u_l)$ for all $o \in \O$ and the evaluation of the transposed derivative multiplied by a data object that is $\frac{\d o}{\d u_i}^T \bar w$ for all $i = 1, \ldots, l(o)$.

\section{Implementing AD for DSL}
The data format for the specification of the language L and the additional derivative information is the most critical point.
The best option would be some kind of pseudo code or real code that can be parsed and used for the generation.
Since this is the first implementation we are using the format of the code generator.
The code generation is done with the gsl library \cite{imatrix2018gsl} which allows a very dynamic mixing of generated code and generation specific logic.
The data format for gsl is xml \cite{bray1997extensible} and therefore the specification for the language $L$ is also done in xml.

A new data object is created with
\begin{lstlisting}[style=xml, caption={Structure definition.},label=lst.struct]
	<structure name="Matrix">
	</structure>
\end{lstlisting}
This will trigger the generation of the type \lstinline|ActiveMatrix<T>|.
Currently the underlying type is just defined as a template but can be specified directly in future implementations.
The class itself contains the storage for the type \lstinline|T| and the identifier.
Furthermore, it provides getter and setter functions for the members as well as some AD specific forwards to the implementation of the global tape from figure \ref{fig.specTape}.
The xml code will also trigger the generation of a \lstinline|MatrixExpression| type which is used by AD to create expression templates in order to improve the recording process.
For an overview about expression templates see \cite{Veldhuizen1995,Aubert2001} and for an AD implementation with expression templates see \cite{Hogan2014FRM, sagebaum2017high}.

The operators are defined via a xml structure:
\begin{lstlisting}[style=xml]
	<function name="<name>" rType="<structure name>">
	  <arg input="{0|1}" type="<structure name>" name="<arg name>">
	    <reverse> ... code ... </reverse>
	  </arg>
	  <primal> ... code ... </primal>
	</function>
\end{lstlisting}
\lstinline[style=xml]|<name>| provides the name of the operator.
\lstinline[style=xml]|<structure name>| has to be the name of the structures defined as shown in listing \ref{lst.struct}.
\lstinline[style=xml]|<arg name>| is the name of an argument.
The element \lstinline[style=xml]|<arg>| can be specified multiple times and the nested \lstinline[style=xml]|<reverse>| element specifies the code for the evaluation of $\frac{\d o}{\d u_i}^T \bar w$ as discussed above.
The \lstinline[style=xml]|<primal>| element specifies the computation of $o(u_1, \ldots, u_l)$.
The \lstinline[style=xml]|<function>| element can either be specified inside a \lstinline[style=xml]|<structure>| element which will generate a member function or in the global xml body which will generate a function.
An example for the multiplication of a matrix and a vector would look like:
\begin{lstlisting}[style=xml]
<function name="mult" rType="Vector">
  <arg input="1" type="Matrix" name="m">
    <reverse> return r_b * v.transpose(); </reverse>
  </arg>
  <arg input="1" type="Vector" name="v">
    <reverse> return m.transpose() * r_b; </reverse>
  </arg>
  <primal> return m * v; </primal>
</function>
\end{lstlisting}

The generated function and expression object is then:
\begin{lstlisting}[style=xml]
template< typename E_m, typename E_v >
struct E_mult_MatVec_AA :
    public VecExpr<E_mult_MatVec_AA<E_m, E_v>> {
  E_m m;
  E_v v;
  typedef typename E_v::RType RType;
  typedef typename E_m::RType A_m;
  typedef typename E_v::RType A_v;

  E_mult_MatVec_AA(const E_m& m, const E_v& v) : m(m), v(v) {}

  static RType computeValue(A_m m, A_v v) {
    return m * v;
  }
  static A_m diff_b_m(A_m m, A_v v,
                      RType r, RType r_b) {
    return r_b * v.transpose();
  }
  static A_v diff_b_v(A_m m, A_v v,
                      RType r, RType r_b) {
    return m.transpose() * r_b;
  }

  RType getValue() const {
    return computeValue(m.getValue(), v.getValue());
  }

  ...
};

template <typename E_m, typename E_v>
E_mult_MatVec_AA<E_m,E_v> mult(
    const MatExpr<E_m>& m,
    const VecExpr<E_v>& v) {
  return E_mult_MatVec_AA<E_m,E_v>(m.cast(), v.cast());
}
\end{lstlisting}

The code in the code sections is included in the generated functions.
Each argument has the type of the original data object.
The return value is provided as an argument with the name \lstinline|r| and for member functions the class itself is provided as an argument with the name \lstinline|t|.
For all involved arguments the corresponding bar value is provided with the same type as the corresponding primal value .
The name is extended by \lstinline|_b|.
The example shows only the generation for the case where both arguments are active (AA).
The generation of the cases with one passive argument (PA and AP) are generated accordingly.

\section{Tests}\label{sec.test}

\subsection{Coupled Burgers equations}
The coupled Burgers equations \cite{biazar2009exact} are chosen as a first test case.
They are used to compare the performance of the new AD tool to some existing AD tools.
This ensures that now major performance killers are used in the implementation.

The coupled Burgers equations
\begin{align}
	u_t + uu_x + vu_y &= \frac{1}{R}(u_{xx} + u_{yy}) \\
	v_t + uv_x + vv_y &= \frac{1}{R}(v_{xx} + v_{yy})
\end{align}
are discretised with an upwind finite difference scheme. The initial conditions are
\begin{align}
	u(x, y, 0) &= x + y \quad (x,y) \in D \\
	v(x, y, 0) &= x - y \quad (x,y) \in D
\end{align}
and the exact solution is after \cite{biazar2009exact},
\begin{align}
	u(x, y, t) &= \frac{x + y - 2xt}{1 - 2t^2} \quad (x,y,t) \in D \times \R\\
	v(x, y, t) &= \frac{x + y - 2xt}{1 - 2t^2} \quad (x,y,t) \in D \times \R \eqdot
\end{align}

The computational domain $D$ is the unit square $D = [0,1] \times [0,1] \subset \R \times \R$ and the boundary conditions are taken from the exact solution.
For the test runs a grid size of 601x601 grid points and 32 discrete time steps are used.
The computations are performed on the Elwetritsch cluster of the TU Kaiserslautern and the test case is evaluated on one node of the cluster which consists of two Intel E5-2640v3 processors.
Two load cases are considered.
For the first case, only one process is run.
For the second, the sequential program is run on each of the 16 cores simultaneously.
This simulates an environment where the memory bandwidth of the node is fully utilized.
The two cases are called `sequential' and `bandwidth limited' in the further analysis.
As a comparison for the DSL AD tool we have chosen CoDiPack \cite{sagebaum2017high} since the basic infrastructure from there is used for the new AD tool.

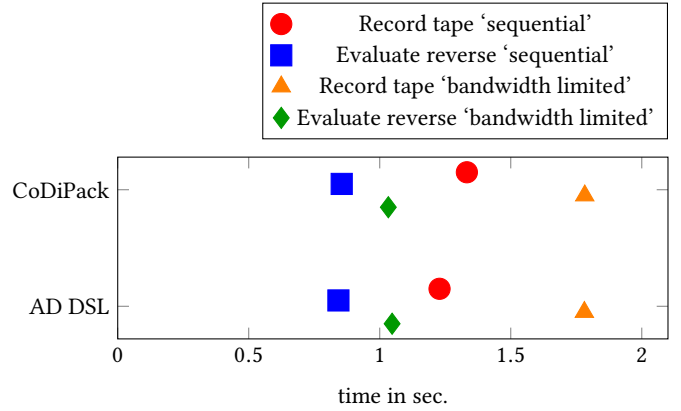
\begin{figure}
	\center
	\begin{tikzpicture}
			\begin{axis}[
				height=4cm,
				width=0.5\textwidth,
				xlabel={time in sec.},
				xmin = 0, xmax = 2.1,
				ytick={-1, -2, -3,  -4, -5, -6},
				yticklabels={CoDiPack , AD DSL},
				legend style={at={(1,1.85)},anchor=north east},
				at={(0,135)}
			]
				\addplot+[symbolDefaults] table[x index=0, y expr=\thisrowno{6}+0.15] \timesBESingleTable;
				\addplot+[symbolDefaults] table[x index=3, y expr=\thisrowno{6}+0.05] \timesBESingleTable;
				\addplot+[symbolDefaults] table[x index=0, y expr=\thisrowno{6}-0.05] \timesBEMultTable;
				\addplot+[symbolDefaults] table[x index=3, y expr=\thisrowno{6}-0.15] \timesBEMultTable;
				\addlegendentry{Record tape `sequential'}
				\addlegendentry{Evaluate reverse `sequential'}
				\addlegendentry{Record tape `bandwidth limited'}
				\addlegendentry{Evaluate reverse `bandwidth limited'}
			\end{axis}
		\end{tikzpicture}	
	\caption{Time comparison for the burgers test case.}
	\label{fig.resultsBurgers}
\end{figure}

The results for the time Measurements are presented in figure \ref{fig.resultsBurgers}.
They show, that the added complexity for the handling of the DSL operators and types, does not increase the required time.
For the recording time in the sequential case the DSL AD tool is even slightly faster than CoDiPack.
The DSL AD tool is currently not as heavily templated as CoDiPack which might provide the compiler with opportunities for optimization.
The memory for both tools is also identical since they store the same data.

\pgfplotstableread{
1.33249e+00 1.44485e-01 5.19815e-01 8.54832e-01 8.51625e-02 2.92667e-01 -1
1.22818e+00 8.15825e-02 5.33118e-01 8.42310e-01 4.40703e-02 1.20140e-01 -2
}\timesBESingleTable
\pgfplotstableread{
1.78240e+00 1.89598e-01 3.25402e-01 1.03274e+00 4.89083e-02 1.21962e-01 -1
1.78088e+00 2.36781e-01 1.25119e-01 1.04722e+00 4.27183e-02 7.60817e-02 -2
}\timesBEMultTable

\subsection{Spline evaluation}

For the DSL specific test case the spline interpolation example from the Vc library \cite{Kretz2015} is chosen.
The interpolation is performed on a $[-1, 1]$ unit square where each dimension of the square is divided into $N$ regions.
The splines for the interpolation in each sub region are third order splines in both dimensions.
The vectorization is done over the nodes of the splines.
They are three dimensional and handled with SIMD vectors of size four.
The example is implemented to run with single precision and therefore the AD tools are also changed to single precision.

The computations are performed on the Elwetritsch cluster of the TU Kaiserslautern and the test case is evaluated on one node of the cluster which consists of two Intel E5-2640v3 processors.
Here we currently consider only the load case where one thread is executed.
This means that the test is not memory bandwidth limited.
Each test consists of the evaluation of $4\cdot 10^6$ samples points.
These points are generated randomly which is also true for the coefficients of the spline interpolation.
The time results are averaged over 20 runs.
For the comparison CoDiPack is again used as a baseline value, in addition the DSL AD tool is once run with the scalar implementation of the interpolation and once run with the vectorized implementation.
All evaluations of the sample points are recorded in one large tape and evaluated accordingly.
This reduces the management overhead for the tape in the provided time measurements.

\begin{figure}
	\center
	\begin{tikzpicture}
			\begin{axis}[
				height=4cm,
				width=0.4\textwidth,
				xlabel={time in sec.},
				xmin = 0, xmax = 5.3,
				ytick={-1, -2, -3,  -4, -5, -6},
				yticklabels={CoDiPack , AD DSL scalar, AD DSL vectorized},
				legend style={at={(1,1.55)},anchor=north east},
				at={(0,135)}
			]
				\addplot+[symbolDefaults] table[x index=0, y expr=\thisrowno{2}+0.15] \timesSplineTable;
				\addplot+[symbolDefaults] table[x index=1, y expr=\thisrowno{2}+0.05] \timesSplineTable;
				\addlegendentry{Record tape `sequential'}
				\addlegendentry{Evaluate reverse `sequential'}
			\end{axis}
		\end{tikzpicture}	
	\caption{Time comparison for the spline test case.}
	\label{fig.resultsSpline}
\end{figure}
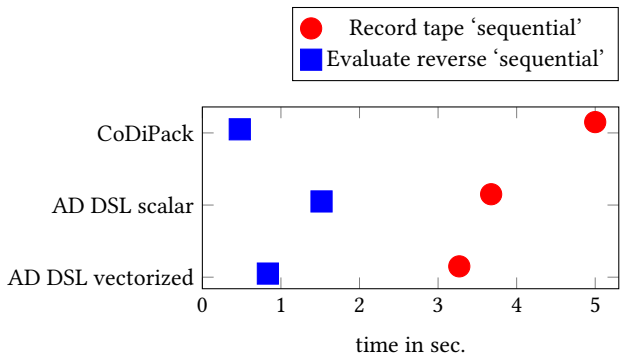

The recording time for CoDiPack is roughly 1.5 seconds slower during the recording of the tape than the AD DSL tool as shown in figure \ref{fig.resultsSpline}.
If the evaluation times are considered then the relation is reversed.
Here, the DSL AD tool is one second slower than CoDiPack.
This discrepancy is in contrast to the Burgers test case and might come from the increased amount of constant values which is introduced by the interpolation nodes of the spline.
CoDiPack can use the constant data directly from the stream since all arguments of the expression have the same type.
This is not possible for the DSL AD tool as it needs to restore the constant data to type specific vectors such that the static evaluation of the expressions can access the data from these vectors.

The comparison of the scalar and vectorized implementation with the DSL AD tool yields the expected results.
The recording time is slightly decreased since less statements are evaluated and stored.
Overall, the general overhead from the tape handling still dominates the recording process.
For the reverse evaluation of the tape the situation is the same.
A greater decrease in time is seen here since the SIMD operations also increase the performance.

For CoDiPack and the scalar DSL AD implementation the memory is the same and around 4.9 Gb.
The vectorized version requires only 3.7 Gb which is a reduction of 24\%.

\pgfplotstableread{
4.9999 0.47626 -1
3.6754 1.5167 -2
3.27 0.83386 -3
}\timesSplineTable

\section{Conclusion}

The paper demonstrates a new approach for the handling of Domain Specific Languages in the context of Algorithmic Differentiation.
Instead of inserting the AD tool into the library of the DSL, a new AD type is generated for every type of the DSL.
The results show that the presented approach is as fast as state of the art AD tools for general applications.
In an applications that uses SIMD operations the performance of the new approach is increased since the SIMD optimizations can still be used.
Furthermore, this leads to a reduced amount of memory.

Since the results are very promising the AD tool for DSLs will be developed further.
The code will be cleaned up and documentation will be written.
It is also necessary to include the handling of multiple output values.
For the test cases in this paper it was not necessary and therefore omitted in the current implementation.
Further steps are then to add additional helper methods for the convenience of the user.

\bibliographystyle{ACM-Reference-Format}
\bibliography{literature}


\begin{thebibliography}{18}


\ifx \showCODEN    \undefined \def \showCODEN     #1{\unskip}     \fi
\ifx \showDOI      \undefined \def \showDOI       #1{#1}\fi
\ifx \showISBNx    \undefined \def \showISBNx     #1{\unskip}     \fi
\ifx \showISBNxiii \undefined \def \showISBNxiii  #1{\unskip}     \fi
\ifx \showISSN     \undefined \def \showISSN      #1{\unskip}     \fi
\ifx \showLCCN     \undefined \def \showLCCN      #1{\unskip}     \fi
\ifx \shownote     \undefined \def \shownote      #1{#1}          \fi
\ifx \showarticletitle \undefined \def \showarticletitle #1{#1}   \fi
\ifx \showURL      \undefined \def \showURL       {\relax}        \fi
\providecommand\bibfield[2]{#2}
\providecommand\bibinfo[2]{#2}
\providecommand\natexlab[1]{#1}
\providecommand\showeprint[2][]{arXiv:#2}

\bibitem[\protect\citeauthoryear{Apostol}{Apostol}{1969}]%
        {apostol1969calculus}
\bibfield{author}{\bibinfo{person}{T.~M. Apostol}.}
  \bibinfo{year}{1969}\natexlab{}.
\newblock \bibinfo{booktitle}{\emph{{Calculus, Volume 2}}}.
\newblock \bibinfo{publisher}{John Wiley \& Sons, Inc.}
\newblock


\bibitem[\protect\citeauthoryear{Aubert, {Di C{\'{e}}sar{\'{e}}}, and
  Pironneau}{Aubert et~al\mbox{.}}{2001}]%
        {Aubert2001}
\bibfield{author}{\bibinfo{person}{P. Aubert}, \bibinfo{person}{N. {Di
  C{\'{e}}sar{\'{e}}}}, {and} \bibinfo{person}{O. Pironneau}.}
  \bibinfo{year}{2001}\natexlab{}.
\newblock \showarticletitle{{Automatic differentiation in C++ using expression
  templates and application to a flow control problem}}.
\newblock \bibinfo{journal}{\emph{Computing and Visualization in Science}}
  \bibinfo{volume}{3}, \bibinfo{number}{4} (\bibinfo{year}{2001}),
  \bibinfo{pages}{197--208}.
\newblock
\showISSN{1432-9360}
\urldef\tempurl%
\url{https://doi.org/10.1007/s007910000048}
\showDOI{\tempurl}


\bibitem[\protect\citeauthoryear{Biazar and Aminikhah}{Biazar and
  Aminikhah}{2009}]%
        {biazar2009exact}
\bibfield{author}{\bibinfo{person}{J. Biazar} {and} \bibinfo{person}{H.
  Aminikhah}.} \bibinfo{year}{2009}\natexlab{}.
\newblock \showarticletitle{{Exact and numerical solutions for non-linear
  Burger’s equation by VIM}}.
\newblock \bibinfo{journal}{\emph{Mathematical and Computer Modelling}}
  \bibinfo{volume}{49}, \bibinfo{number}{7} (\bibinfo{year}{2009}),
  \bibinfo{pages}{1394--1400}.
\newblock


\bibitem[\protect\citeauthoryear{Bischof, Carle, Khademi, and Mauer}{Bischof
  et~al\mbox{.}}{1996}]%
        {Bischof1996AAD}
\bibfield{author}{\bibinfo{person}{C.~H. Bischof}, \bibinfo{person}{A. Carle},
  \bibinfo{person}{P. Khademi}, {and} \bibinfo{person}{A. Mauer}.}
  \bibinfo{year}{1996}\natexlab{}.
\newblock \showarticletitle{{ADIFOR} 2.0: Automatic Differentiation of
  {F}ortran 77 Programs}.
\newblock \bibinfo{journal}{\emph{IEEE Computational Science \& Engineering}}
  \bibinfo{volume}{3}, \bibinfo{number}{3} (\bibinfo{year}{1996}),
  \bibinfo{pages}{18--32}.
\newblock


\bibitem[\protect\citeauthoryear{Bray, Paoli, Sperberg-McQueen, Maler, and
  Yergeau}{Bray et~al\mbox{.}}{1997}]%
        {bray1997extensible}
\bibfield{author}{\bibinfo{person}{T. Bray}, \bibinfo{person}{J. Paoli},
  \bibinfo{person}{C.~M. Sperberg-McQueen}, \bibinfo{person}{E. Maler}, {and}
  \bibinfo{person}{F. Yergeau}.} \bibinfo{year}{1997}\natexlab{}.
\newblock \showarticletitle{Extensible markup language (XML).}
\newblock \bibinfo{journal}{\emph{World Wide Web Journal}} \bibinfo{volume}{2},
  \bibinfo{number}{4} (\bibinfo{year}{1997}), \bibinfo{pages}{27--66}.
\newblock


\bibitem[\protect\citeauthoryear{Dunford and Schwartz}{Dunford and
  Schwartz}{1958}]%
        {dunford1958linear}
\bibfield{author}{\bibinfo{person}{N. Dunford} {and} \bibinfo{person}{J.T.
  Schwartz}.} \bibinfo{year}{1958}\natexlab{}.
\newblock \bibinfo{booktitle}{\emph{{Linear Operators: General theory}}}.
\newblock \bibinfo{publisher}{Interscience Publishers}.
\newblock
\showLCCN{57010545}
\urldef\tempurl%
\url{https://books.google.de/books?id=i2EPAQAAMAAJ}
\showURL{%
\tempurl}


\bibitem[\protect\citeauthoryear{Firasta, Buxton, Jinbo, Nasri, and
  Kuo}{Firasta et~al\mbox{.}}{2008}]%
        {firasta2008intel}
\bibfield{author}{\bibinfo{person}{N. Firasta}, \bibinfo{person}{M. Buxton},
  \bibinfo{person}{P. Jinbo}, \bibinfo{person}{K. Nasri}, {and}
  \bibinfo{person}{S. Kuo}.} \bibinfo{year}{2008}\natexlab{}.
\newblock \showarticletitle{{Intel AVX: New frontiers in performance
  improvements and energy efficiency}}.
\newblock \bibinfo{journal}{\emph{Intel white paper}}  \bibinfo{volume}{19}
  (\bibinfo{year}{2008}), \bibinfo{pages}{20}.
\newblock


\bibitem[\protect\citeauthoryear{Griewank and Walther}{Griewank and
  Walther}{2008}]%
        {grie08}
\bibfield{author}{\bibinfo{person}{A. Griewank} {and} \bibinfo{person}{A.
  Walther}.} \bibinfo{year}{2008}\natexlab{}.
\newblock \bibinfo{booktitle}{\emph{{Evaluating Derivatives, second edition}}}.
\newblock \bibinfo{publisher}{SIAM}.
\newblock
\showISBNx{978-0-898716-59-7}


\bibitem[\protect\citeauthoryear{Hasco{\"e}t and Pascual}{Hasco{\"e}t and
  Pascual}{2013}]%
        {TapenadeRef13}
\bibfield{author}{\bibinfo{person}{L. Hasco{\"e}t} {and} \bibinfo{person}{V.
  Pascual}.} \bibinfo{year}{2013}\natexlab{}.
\newblock \showarticletitle{{The Tapenade Automatic Differentiation tool:
  Principles, model, and specification}}.
\newblock \bibinfo{journal}{\emph{{ACM} {TOMS}}} \bibinfo{volume}{39},
  \bibinfo{number}{3} (\bibinfo{year}{2013}).
\newblock
\urldef\tempurl%
\url{http://dx.doi.org/10.1145/2450153.2450158}
\showURL{%
\tempurl}


\bibitem[\protect\citeauthoryear{Hogan}{Hogan}{2014}]%
        {Hogan2014FRM}
\bibfield{author}{\bibinfo{person}{R.~J. Hogan}.}
  \bibinfo{year}{2014}\natexlab{}.
\newblock \showarticletitle{{Fast reverse-mode Automatic Differentiation using
  expression templates in C++}}.
\newblock \bibinfo{journal}{\emph{ACM TOMS}} \bibinfo{volume}{40},
  \bibinfo{number}{4} (\bibinfo{year}{2014}), \bibinfo{pages}{26:1--26:24}.
\newblock
\urldef\tempurl%
\url{http://doi.acm.org/10.1145/2560359}
\showURL{%
\tempurl}


\bibitem[\protect\citeauthoryear{iMatix Corporation}{iMatix
  Corporation}{2018}]%
        {imatrix2018gsl}
\bibfield{author}{\bibinfo{person}{iMatix Corporation}.}
  \bibinfo{year}{2018}\natexlab{}.
\newblock \bibinfo{title}{{GSL}}.
\newblock \bibinfo{howpublished}{\url{https://github.com/imatix/gsl}}.
  (\bibinfo{year}{2018}).
\newblock
\newblock
\shownote{[Online; accessed 10-January-2018].}


\bibitem[\protect\citeauthoryear{Kretz}{Kretz}{2015}]%
        {Kretz2015}
\bibfield{author}{\bibinfo{person}{M. Kretz}.} \bibinfo{year}{2015}\natexlab{}.
\newblock \emph{\bibinfo{title}{{Extending C++ for explicit data-parallel
  programming via SIMD vector types}}}.
\newblock \bibinfo{thesistype}{Ph.D. Dissertation}.
\newblock


\bibitem[\protect\citeauthoryear{Naumann}{Naumann}{2012}]%
        {naumann2012art}
\bibfield{author}{\bibinfo{person}{U. Naumann}.}
  \bibinfo{year}{2012}\natexlab{}.
\newblock \bibinfo{booktitle}{\emph{{The Art of Differentiating Computer
  Programs: An Introduction to Algorithmic Differentiation}}}.
\newblock \bibinfo{publisher}{SIAM}.
\newblock
\showISBNx{9781611972061}
\showLCCN{2011032262}
\urldef\tempurl%
\url{https://books.google.de/books?id=VpheBMaDf4kC}
\showURL{%
\tempurl}


\bibitem[\protect\citeauthoryear{Sagebaum, Albring, and Gauger}{Sagebaum
  et~al\mbox{.}}{2017}]%
        {sagebaum2017high}
\bibfield{author}{\bibinfo{person}{M. Sagebaum}, \bibinfo{person}{T. Albring},
  {and} \bibinfo{person}{N.~R. Gauger}.} \bibinfo{year}{2017}\natexlab{}.
\newblock \showarticletitle{{High-Performance Derivative Computations using
  CoDiPack}}.
\newblock \bibinfo{journal}{\emph{arXiv preprint arXiv:1709.07229}}
  (\bibinfo{year}{2017}).
\newblock
\urldef\tempurl%
\url{https://arxiv.org/abs/1709.07229}
\showURL{%
\tempurl}


\bibitem[\protect\citeauthoryear{Sagebaum, Albring, and Gauger}{Sagebaum
  et~al\mbox{.}}{2018}]%
        {sagebaum2018Expr}
\bibfield{author}{\bibinfo{person}{M. Sagebaum}, \bibinfo{person}{T. Albring},
  {and} \bibinfo{person}{N.~R. Gauger}.} \bibinfo{year}{2018}\natexlab{}.
\newblock \showarticletitle{{Expression templates for primal value taping in
  the reverse mode of Algorithmic Differentiation}}.
\newblock \bibinfo{journal}{\emph{submitted to Optimization Methods \&
  Software}} (\bibinfo{year}{2018}).
\newblock


\bibitem[\protect\citeauthoryear{Van~Deursen, Klint, and Visser}{Van~Deursen
  et~al\mbox{.}}{2000}]%
        {van2000domain}
\bibfield{author}{\bibinfo{person}{A. Van~Deursen}, \bibinfo{person}{P. Klint},
  {and} \bibinfo{person}{J. Visser}.} \bibinfo{year}{2000}\natexlab{}.
\newblock \showarticletitle{{Domain-specific languages: An annotated
  bibliography}}.
\newblock \bibinfo{journal}{\emph{ACM Sigplan Notices}} \bibinfo{volume}{35},
  \bibinfo{number}{6} (\bibinfo{year}{2000}), \bibinfo{pages}{26--36}.
\newblock


\bibitem[\protect\citeauthoryear{Veldhuizen}{Veldhuizen}{1995}]%
        {Veldhuizen1995}
\bibfield{author}{\bibinfo{person}{T. Veldhuizen}.}
  \bibinfo{year}{1995}\natexlab{}.
\newblock \showarticletitle{{Expression templates}}.
\newblock \bibinfo{journal}{\emph{C++ Report}}  \bibinfo{volume}{7}
  (\bibinfo{year}{1995}), \bibinfo{pages}{26--31}.
\newblock
\showISSN{1040-6042}


\bibitem[\protect\citeauthoryear{Walther and Griewank}{Walther and
  Griewank}{2009}]%
        {walther2009getting}
\bibfield{author}{\bibinfo{person}{A. Walther} {and} \bibinfo{person}{A.
  Griewank}.} \bibinfo{year}{2009}\natexlab{}.
\newblock \showarticletitle{{Getting started with ADOL-c.}}. In
  \bibinfo{booktitle}{\emph{Combinatorial scientific computing}}.
  \bibinfo{pages}{181--202}.
\newblock


\end{thebibliography}

\end{document}